\documentstyle[art12,epsfig]{article}
\newcommand {\vp}[1] {\vec{p}_{#1}}
\newcommand {\pom} {{ I \hspace {-2.6pt} P}}
\begin{document}

%
%
%
\centerline{\Large \bf Diffraction in Deep Inelastic Scattering:}
\centerline{\Large \bf the hadronic nature of Quarks}
\bigskip

\flushleft{ L. Dick, \, CERN, Geneva, Switzerland}
\flushleft{V. Karapetian, \, CERN, Geneva, Switzerland}
\flushleft{R. Barni, \, Dip. di Fisica, Univ. degli Studi, and INFN, Sezione 
di Milano, via~Celoria,~16 \, I-20133 Milano, Italy}
\flushleft{G. Preparata, \, Dip. di Fisica, Univ. degli Studi, 
and INFN, Sezione di Milano, 
via~Celoria,~16 \, I-20133 Milano, Italy}

\vspace {2 truecm}

\centerline{\Large \bf Abstract}
\bigskip
We analyse the ``diffractive'' events reported by the HERA groups within 
a theoretical framework (Anisotropic Chromo Dynamics, ACD) that 
attributes to the fundamental quanta of QCD (Quarks and Gluons) a 
completely hadronic behaviour except, of course, for the Quark ``point-like''
coupling to the electroweak fields.\par
The remarkable success of our calculation, free of adjustable 
parameters, highlights the fallacy of considering 
``short-distance'' physics as autonomous (indeed orthogonal) from 
the long-distance, colour confining hadronic physics.
\vfill
\flushleft{ \bf MITH-96/4}
%
%
\newpage

\section {Introduction}

One of the most relevant and surprising aspects of the physics produced 
by HERA in the last four years is the discovery of a large number of events 
whose final states are characterised by large rapidity gaps~\cite{disc}, 
of the kind one observes in diffractive events in purely 
hadronic interactions~\cite{gou}.
This occurrence is all the more surprising since ``Diffraction'', or in 
general the physics of the Pomeron, is considered to be the genuine 
manifestation of the dynamics of QCD at large distances, or at small 
transverse momenta, where perturbation theory (PQCD) has absolutely 
nothing to say, the main actors being the Hadrons, the permanent 
prisons of Quarks and Gluons. On the other hand Deep Inelastic 
Scattering (DIS) is generally considered the true realm of QCD at short 
space-time distances where Asymptotic Freedom (AF) should allow us, 
should it not?, to compute the basic inclusive cross-section through the 
simple Feynman rules of PQCD.\par
In hindsight a number of PQCD mechanisms~\cite{bu,zac} have been invoked in 
order to save the general expectations based on AF, but the puzzle 
that stands in the way of the natural philosopher is to understand why a 
typical hadronic behaviour, completely extraneous to PQCD, can indeed be 
mimicked in a world thoroughly governed by this simple (unconfined) 
realization of QCD.\par
Almost a quarter of a century ago one of us (G.P.) found that the 
fundamental aspects of the just discovered Bjorken scaling behaviour 
could be easily and naturally accounted for by assuming that Quarks (and 
later Gluons) would behave at any space-time scale like hadronic 
particles~\cite{mqm}, with their Regge behaviours that were known since 
long to well describe high energy interactions~\cite{reg}. 
Naturally in the Massive Quark Model (MQM)~\cite{mqm} no attempt 
was made to derive ``Reggeism'' from any 
basic Lagrangian, the only firm Ansatz being that Regge behaviour, and 
in particular the Pomeron and Diffraction, should be properties of the 
Quarks as well. It should be perfectly clear that such an attitude was 
then (and is now) completely at variance with the beliefs and the 
expectations of the vast majority of the particle physicists; however it 
is a fact that the recent observations at HERA do agree with 
those distant attempts, while they have posed and continue to pose some 
non-trivial problems to the theorists of PQCD.\par
While along the paths indicated by AF PQCD has 
developed into an extremely powerful and flexible means to deal with 
DIS physics, the MQM has gone through the much more uncertain steps that 
have led it to Anisotropic Chromo Dynamics (ACD)~\cite{acd}, the  
``effective'' Lagrangian which governs the dynamics of QCD over a 
peculiar confining vacuum (the Chromo Magnetic Liquid, CML~\cite{cml}) 
that has definite chances to well approximate the true QCD vacuum. 
Recently one has been able to find within ACD a simple and 
satisfactory approach to the Pomeron and Diffraction~\cite{pgr} that 
embodies in a physically transparent way the pioneering 
approach of Low and Nussinov~\cite{low}. 

\section{Diffractive Deep Inelastic Scattering}

The basic idea is that when the 
Quarks of two Hadrons collide at high energy (see Fig.1) and interact by 
the fundamental colour-exchange potential, the Hadrons (the 
colour-singlets) of the final state comprise two different (in our case) 
$Q \bar Q$-pairs whose masses are of the order of the CM energies,  
which then fill the final states with the products of their decay~\cite{fs}. 
A detailed calculation~\cite{pgr} shows that the diagrams in Fig.1 
correctly reproduce the main characteristic of the Pomeron.
As for single Diffraction in DIS the relevant diagrams are reported in 
Fig.2, where it should be noticed that the deep inelastic photon 
couples to the Quarks through a ``point-like wave function''.\par
The virtual photon-proton diffractive cross section is given by a
tedious but straightforward calculation of the Feynman-like diagrams of 
Fig.2, whose only ``unorthodox'' elements are: 
\begin{itemize}
\item[(i)] the proton wave-function that projects the initial hadronic state 
(the proton) onto its constituent quark state which in the high energy 
limit $(W_{\gamma^* p} \rightarrow \infty , x_{Bj} \rightarrow 0)$ can be 
taken as:
\begin{equation}
|\vec{p}> = \int d^3\vp{1} d^3\vp{2} \sqrt{\frac {(2\pi)^3 2 E_p} {(2\pi)^9 
2E_1 2E_2 2E_3}}  \: \phi (\vp{1},\vp{2}) \, 
| \vp{1},\vp{2},\vp{}-\vp{1}-\vp{2}>
\end{equation}
with the normalisation 
\begin{equation}
\int d^3\vp{1} d^3\vp{2} \, \left| \phi (\vp{1},\vp{2}) \right|^2=1 ,
\end{equation}
and the factorisation property 
$(\vp{i}=x_i \vp{}+\vp{\perp i})$, 
$|\vp{\perp i}| \ll |\vp{}|$
\begin{equation}
\phi= \phi_L (x_1,x_2) \cdot \phi_T (\vp{\perp1}, \vp{\perp2}) ,
\end{equation}
which allows the complete separation of the trivial longitudinal dynamics
from the transverse one. We parameterise 
$\phi_T(\vp{1\perp},\vp{2\perp})$ 
as a Gaussian distribution~\cite{pgr}
\begin{equation}
\phi_T(\vp{1\perp},\vp{2\perp})=
\sqrt{12} \frac{b}{\pi} exp[-b (p^2_{\perp 1} + p^2_{\perp 2} 
+(\vp{\perp1} + \vp{\perp2})^2)]
\end{equation}
with 
$b=17$ GeV$^{-2}$, 
as determined by the slope of the forward peak in the pp elastic 
cross section. 
\item[(ii)] the diffractive final state comprises two  Fire String (FS) 
states~\cite{fs}, whose decay products are the Hadrons of the final states. 
Please note  that in the inclusive cross-section Quark-Hadron 
``duality''~\cite{dua} allows us to integrate over the quark degrees 
of freedom only; 
\item[(iii)] the ACD gluon propagator~\cite{acd} , whose form can be taken as:
\begin{equation}
G^{ab}_{\mu \nu}(q) = \delta^{ab} g_{\mu \nu} 8 \pi \mu^2 
\frac{3\alpha^2-|q^2|}{(\alpha^2+|q^2|)^3} = \delta^{ab} g_{\mu \nu} 
8 \pi \mu^2 V(q^2), 
\end{equation}
giving rise to an interquark effective potential 
$\tilde{V}(r) = \mu^2 r e^{-\alpha r}$,
where the two parameters $\mu$, the effective string tension, and the 
``longitudinal size'' of the ``chromomagnetic needle'' 
$\alpha^{-1}$, 
are fixed by the analysis of the hadronic spectrum~\cite{sco} 
as ($m_g$ denotes the gluon mass)
\begin{equation}
\mu = 0.475\, {\rm GeV}, \;\;\;
\alpha = 2\sqrt{2}m_g = 0.40 \, {\rm GeV}. 
\end{equation}
\end{itemize}
The result of the calculation can be expressed in terms of a diffractive 
structure function 
$F_2^D$: 
\begin{equation}
\left.  \frac {{d\sigma}^{\gamma^* p \rightarrow X p}}{d\eta_{\pom} dt}
\right |_{diff} = \frac {4 \pi^2 \alpha_{QED}} {Q^2 (1-x_{Bj})} 
 \: \frac {dF^D_2}{d\eta_{\pom} dt} (x_{Bj}, Q^2, \eta_{\pom}, t)
\end{equation}
\begin{equation}
\eta_{\pom}= \frac {M^2_X+Q^2-t}{W^2+Q^2-M^2_p} = \frac{x_{Bj}}{\beta} 
\sim 1-x_F , \, \, \, \, \, \beta=\frac{Q^2}{M^2_X+Q^2-t}
\end{equation}
where ($x_F$ is the Feynman variable of the final state proton, $M_X$ is the
invariant hadronic mass, $t$ is the invariant momentum transferred to the final
state proton). $F^D_2$ is given by the expression
\begin{equation}
\frac {dF^D_2}{d\eta_{\pom} dt}= \frac {N}{\eta_{\pom}} I(t)^2 
J(x_{Bj},Q^2) (1-x_{Bj}),
\end{equation}
where the normalisation contains the couplings, the colour factor, the 
sum over active quark flavour and the other numerical factors,
\begin{equation}
N = \sum_{f} Q^2_f {\cal{C}}_F \cdot \left( \frac{\mu}{\alpha} 
\right)^{12} \cdot \frac {96}{\pi^9 \alpha^2} \sim 1.77 \cdot 10^{-2} \: 
{\rm GeV}^{-2}
\end{equation}
\begin{equation}
I(t)=\int d^2\vec{w} \,  V(\vec{w}-\vec{x}_0) V(\vec{w}+\vec{x}_0)
(e^{-\delta x^2_0} - e^{-\delta w^2})
\end{equation}
with $ \vec{x_0}=\frac {\vp{\perp}}{2\alpha}$ the transverse momentum of 
the final state proton, and $\delta=b \alpha^2 \simeq 3$.
Furthermore
\begin{equation}
J(x,Q^2)= \int d^2\vec{\xi} d^2\vec{z} \:  z^2 \, {\rm ln} \frac 
{W_{\gamma^* p}^2}{z^2} V(\vec{z})^2 [ \phi_{\gamma}(\vec{\xi},0) - 
\phi_{\gamma}(\vec{\xi},\vec{z}) ]
\end{equation}
where the ``photon wave-function'' 
$\phi_{\gamma} (\vec{\xi},\vec{z})$ 
is given by 
$(A=\frac {Q^2}{{\alpha}^2})$:
\begin{equation}
\phi_{\gamma} (\vec{\xi},\vec{z}) = \int^{1}_{0} d\tau \, \frac 
{\tau^2 \vec{\xi} \cdot (\vec{\xi}+\vec{z}) A} 
{[\tau(1-\tau)A+\vec{\xi}^2][\tau(1-\tau)A + (\vec{\xi}+\vec{z})^2]}.
\end{equation}
A few observations on Eqs. (5) to (9) are in order: 
\begin{itemize}
\item[(a)] the energy dependence is logarithmic, thus no ``unitarization''
is necessary to obey the Froissart bound~\cite{fro};
\item[(b)] in the limited kinematic range now accessible to the 
experiments such logarithmic behaviour can well mimic a power law, 
which is favoured by the experimental fits;
\item[(c)] the factorisation of the diffractive structure function as 
\begin{equation}
F_2^D(\eta_{\pom},\beta,Q^2) \sim \Phi(\eta_{\pom}) \cdot F_2^{\pom}(\beta,Q^2) 
\end{equation}
is broken due to the $t$-integration of equation (11) and to the 
contribution of diagrams subleading respect to that of Fig.~2, 
however, as one can see from 
Fig.~\ref{Fig:fgz},
%
%
%
in the experimentally accessible region our result is effectively 
factorized; with factorizable form the $F_2^{\pom}(\beta,Q^2)$ 
can be associated with Pomeron structure function, as it is discussed 
in the pioneer work~\cite{is}.
\item[(d)] when 
$\beta \rightarrow 0$, 
corresponding to very large diffractive masses, the typical HERA energies 
($W_{\gamma^* p} \simeq 100\,$ GeV) 
imply that 
$\eta_{\pom}$ 
becomes rather sizeable, thus requiring the introduction of sub-leading 
Regge trajectories, like 
$\omega,f_2,...,\pi,$ 
as suggested by purely hadronic data~\cite{isr,gou}, that we have neglected 
in our calculation. Since such contributions can alter the form of the flux 
function $\Phi(\eta_{\pom})$, the extraction of $F_2^{\pom}(\beta,Q^2)$ 
can be more safely performed by integrating $F_2^D$ in each bin of the 
$(\beta,Q^2)$-plane, rather than fitting the $F_2^D$-function with a 
universal power law over a fixed $\eta_{\pom}$ interval which is then 
used in the determination of $F_2^{\pom}$, as is commonly done by the HERA 
Collaborations~\cite{zeus,h1}.
\end{itemize}

\section {Comparison with experimental data}

Our calculation of the triple differential structure function 
$F_2^D(\eta_{\pom},\beta,Q^2)$ is compared with the experimental data 
from the ZEUS Collaboration~\cite{zeus} in 
Fig.~\ref{Fig:fgz} 
and from the H1 Collaboration~\cite{h1} in
Fig.~\ref{Fig:fgh}. 
%
%
%
One notes good agreement with observations for all accessible values of
$\beta$,$Q^2$ and $\eta_{\pom}$, leading to overall statistical 
value $\chi^2$/d.f. = 67.4/96 being compared with data.
In order to see the consistency of both H1 and ZEUS  data
we performed a new power-law fit to the combined set of data (shown in  
double-log scale in 
Fig.~\ref{Fig:fgz} 
and 
Fig.~\ref{Fig:fgh}
%
%
%
by straight line dependence),which yielded the universal value of the
Pomeron intercept $\alpha_{\pom}(0)$ = 1.069 $\pm$ 0.030 
($\chi^2$/d.f. = 33.33/47 [H1], 4.05/23 [ZEUS] assuming full errors in 
the fit as statistical and systematic errors added in quadrature).
Within experimental errors, this value of $\alpha_{\pom}(0)$ is 
consistent with a soft supercritical Pomeron~\cite{dl} and is correctly 
reproduced by our model.\par
Our prediction for double differential structure functions  
$F_2^D(\beta,Q^2)$ and $F_2^D(x_{Bj},Q^2)$ is reported for different values
of $Q^2$ in 
Fig.~\ref{Fig:beta} 
%
%
%
and compared with data points from both ZEUS and 
H1~\cite{zeus,h1,h11}. Again good agreement is to be noted. 
In particular a smooth $Q^2$-dependence is naturally reproduced by
the model. To calculate the $\beta$-distributions, the underlying
$\eta_{\pom}$-integration is performed for two different intervals
defined by HERA Collaborations ~\cite{zeus,h1}, and we find that the 
discrepancies in the values of the Pomeron structure function 
measured by H1 and ZEUS does depend on such difference. 
Indeed, as one can see in 
Fig.~\ref{Fig:beta}, 
%
%
%
our model favours a small $\beta$-dependence in the Pomeron structure
function. The present (1993) data are statistically compatible 
with both the model of Ref.~\cite{zeus} and our predicted flat 
$\beta$-dependence. With more precise data it will certainly be possible
to distinguish between the two approaches. But in this case we must be more 
careful in data analysis because the unfolding procedure
(to calculate the acceptance and compensate kinematic cuts effects)
based on Monte Carlo simulations~\cite{zeus,h1,h11} can give 
undesirable correlations between the model to be tested and the corrected 
data~\cite{zeus,h1}, since $\beta$-distribution affects, in
particular, the mass spectrum of hadronic final states. Clearly, 
a better comparison of our model could be made by processing with it
the raw data.\par
Finally, our expectation for the total contribution of Single Diffractive
Dissociation (SDD) process is given in the last four pictures in
Fig.~\ref{Fig:beta},
%
%
%
in terms of scaling variable $x_{Bj}$ and for different values of $Q^2$.
The relevant upper limit for the $\eta_{\pom}$-integration  
must be chosen about 0.15, in accordance with pure hadronic 
data~\cite{isr,gou}. Note, that the decrease of $F_2^D$ to zero at
$x_{Bj} = 10^{-2}$ is due to the special kinematic and experimental
conditions at HERA. When these limitations are removed it is seen
that the real contribution of SDD to DIS is closer to 
20\% than to 10\%, as claimed by the HERA groups.
This discrepancy is due to their procedure for selecting large
rapidity gap events and to the far from ideal acceptance of their setup
for diffractive physics.

\section{Conclusion}

To conclude this letter, we may state with confidence that a sound, 
realistic and conceptually simple explanation has emerged from ACD of the 
complex features of Diffraction in DIS discovered by HERA. The main 
message of this paper, we believe, is the further proof of the correctness
of the idea that deep inelastic phenomena fully exhibit the basic features 
of high energy hadronic dynamics, albeit at the level of the fundamental 
quarks instead of the composite hadrons. On the other hand the hadronic 
nature of the quarks that is corroborated by this work is, to our mind, a 
further stumbling block on the road of PQCD, which looks at the deep 
inelastic world as completely different and separated (by the scale 
$\Lambda_{QCD}$) from the world of long-distance (low $p_T$) phenomena.
Finally we notice that the same ideas can be employed to compute the 
non-diffractive low $x_{Bj}$ part of the structure function, which 
shall be reported in a future publication.

\subsection*{Acknowledgement}

One of us (R.B.) is pleased to recognise the help from P. G. Ratcliffe
and T. Sanvito in the preliminary stage of the work.

\newpage

\centerline{\bf Captions}
\bigskip

{\bf Fig.~1 }
\medskip

The amplitude dominating the total cross-section for 
meson-meson scattering at high energy (all the diagrams obtained by 
permutations of the quark-gluon vertices should be summed).
\bigskip

{\bf Fig.~2}
\medskip

The amplitude dominating the single diffractive cross-section 
for DIS at high energy $W_{\gamma^* p}$ and large diffracted mass $M^2_X$ 
(the so called Triple-Regge regime [12]).
\bigskip

{\bf Fig.~3}
\medskip

The triple differential structure function $F_2^D$ of ZEUS is compared 
with our calculation (broken line), first passing through all the 
experimental cuts (dotted line) and then relaxing the cuts, but in the 
integration region defined by ZEUS (full line) and with a power-law fit 
(dashed line). The almost negligible contribution of non-factorisable
diagrams is also shown (dot-dashed line).
Note that the $15\%$  contribution due to double diffraction
has been subtracted from data [18].
\bigskip

{\bf Fig.~4}
\medskip

The triple differential structure function $F_2^D$ of H1 is compared 
with our calculation (broken line), first passing through all the 
experimental cuts (dotted line) and then relaxing the cuts, but in the 
integration region defined by H1 (full line) and with a power-law fit 
(dashed line). Note that the $15\%$  contribution due to double 
diffraction has been subtracted from data [18].
\bigskip

{\bf Fig.~5}
\medskip

The double differential structure function $F_2^D$ of ZEUS and H1
in terms of  $\beta$ and $x_{Bj}$, predicted by the model (full line) 
and compared with the experimental data. The black points on 
$\beta$-distribution correspond to $\alpha_{\pom}(0)$ = 1.069, obtained 
in our power-law fit to combined set of HERA data. 
The black and white points on $x_{Bj}$-distribution are observations using 
two different analysis methods [21], the vertical lines are pure 
kinematic cuts. Note that the $15\%$  contribution
due to double diffraction has been subtracted from data [18]. 
The last four pictures show the expected contribution of 
single diffraction dissociation, integrated with the relaxed upper 
limit on $\eta_{\pom}$ (see text above).

%
%
\newpage
\begin{figure}[htbp]
\epsfig{file=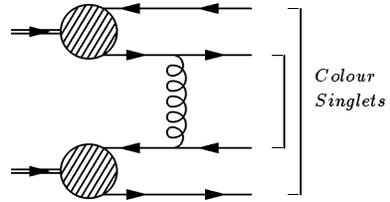,height=21.cm,width=15.5cm}
\caption{The amplitude dominating the total cross-section for 
meson-meson scattering at high energy (all the diagrams obtained by 
permutations of the quark-gluon vertices should be summed).}  
\label{Fig:diatot}
\end{figure}

\newpage
\begin{figure}[htbp]
\epsfig{file=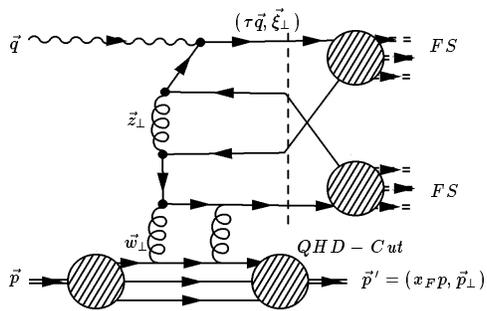,height=21cm,width=15.5cm}
\caption{The amplitude dominating the single diffractive cross-section 
for DIS at high energy $W_{\gamma^* p}$ and large diffracted mass $M^2_X$ 
(the so called Triple-Regge regime [12]).}
\label{Fig:diadiff}
\end{figure}
%
\newpage
\begin{figure}[p]
\epsfig{file=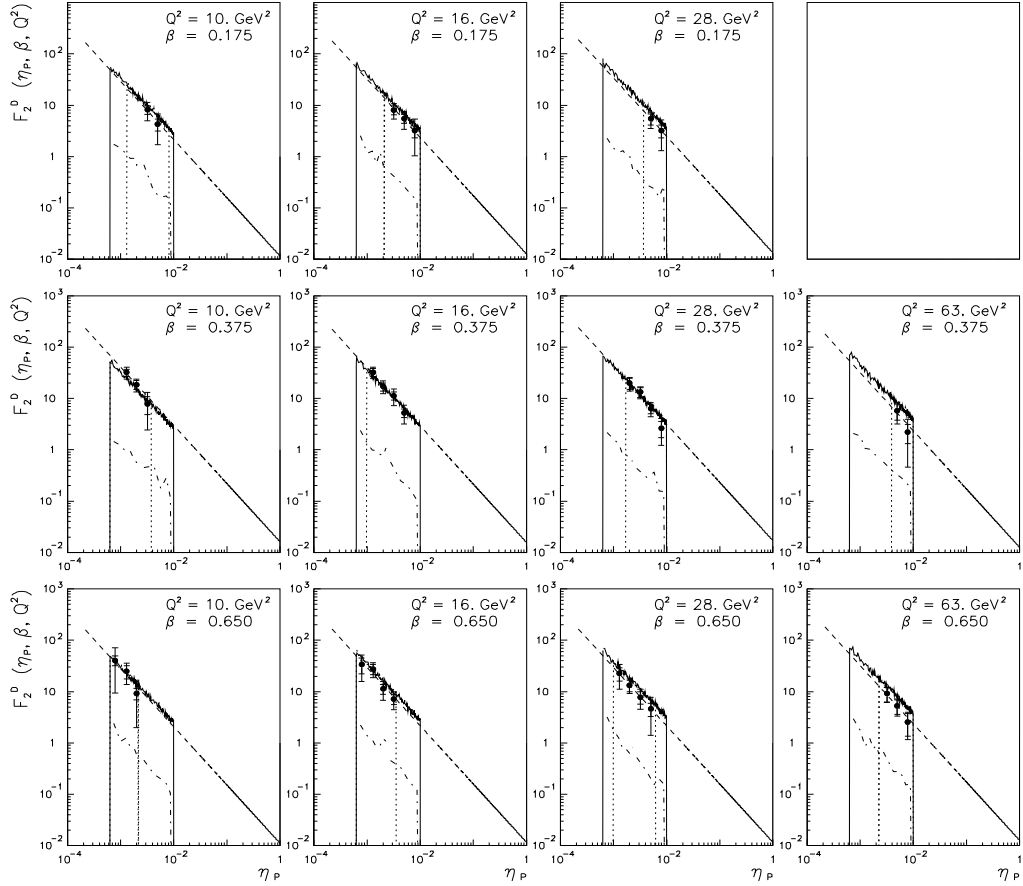,height=19.cm,width=15.5cm}
\caption {The triple differential structure 
function $F_2^D$ of ZEUS is compared with our calculation (broken line), 
first passing through all the experimental cuts (dotted line) and then 
relaxing the cuts, but in the integration region defined by ZEUS
(full line) and with a power-law fit (dashed line).
The almost negligible contribution of non-factorisable
diagrams is also shown (dot-dashed line).
Note that the $15\%$  contribution due to double diffraction
has been subtracted from data [18].}
\label{Fig:fgz}
\end{figure}

\newpage
\begin{figure}[p]
\epsfig{file=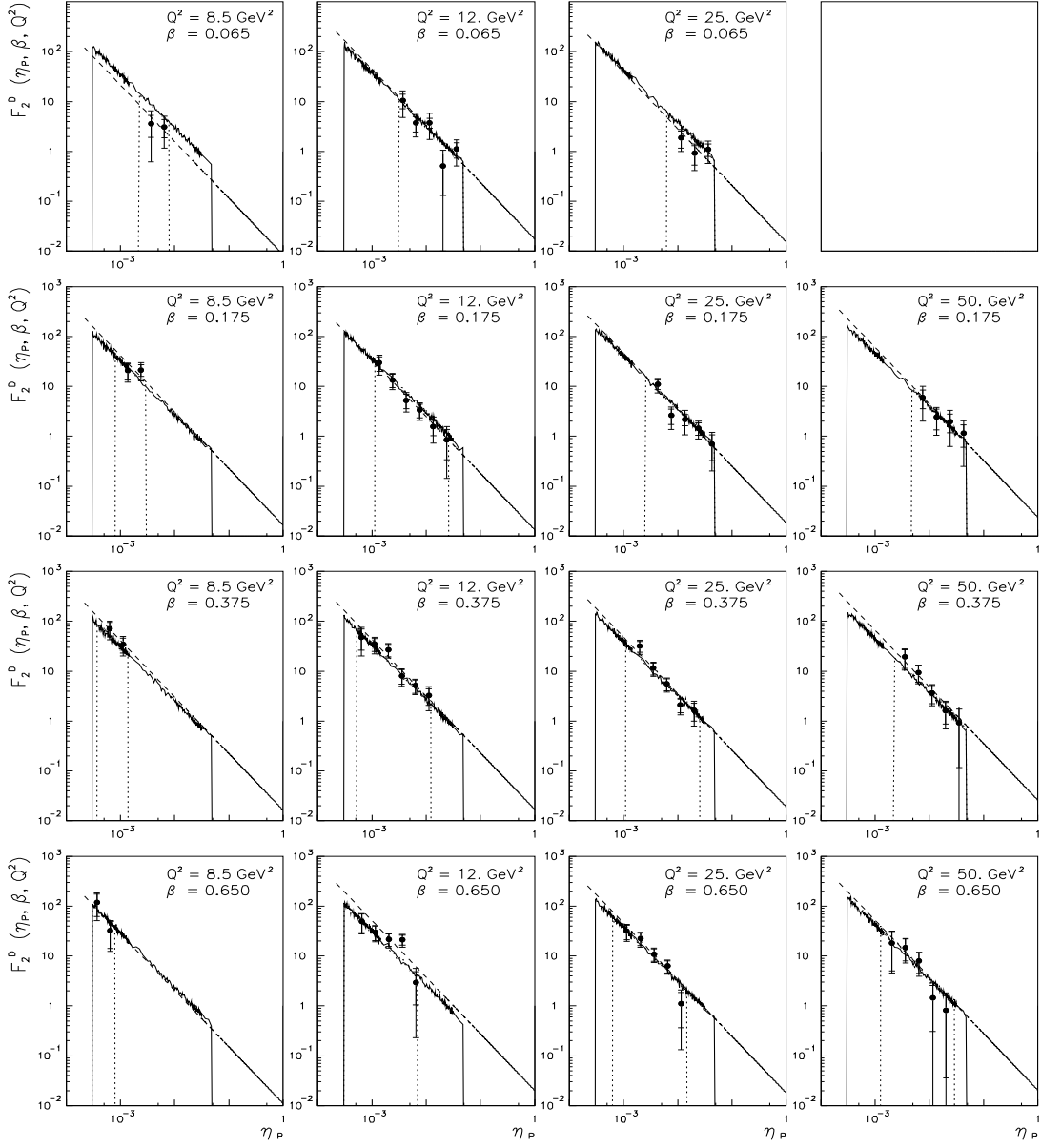,height=19.5cm,width=15.5cm}
\caption 
{The triple differential structure function $F_2^D$ of H1 is compared 
with our calculation (broken line), first passing through all the 
experimental cuts (dotted line) and then relaxing the cuts, but in the 
integration region defined by H1 (full line) and with a power-law fit 
(dashed line). Note that the $15\%$  contribution due to double diffraction
has been subtracted from data [18].}
\label{Fig:fgh}
\end{figure}

\newpage
\begin{figure}[htbp]
\epsfig{file=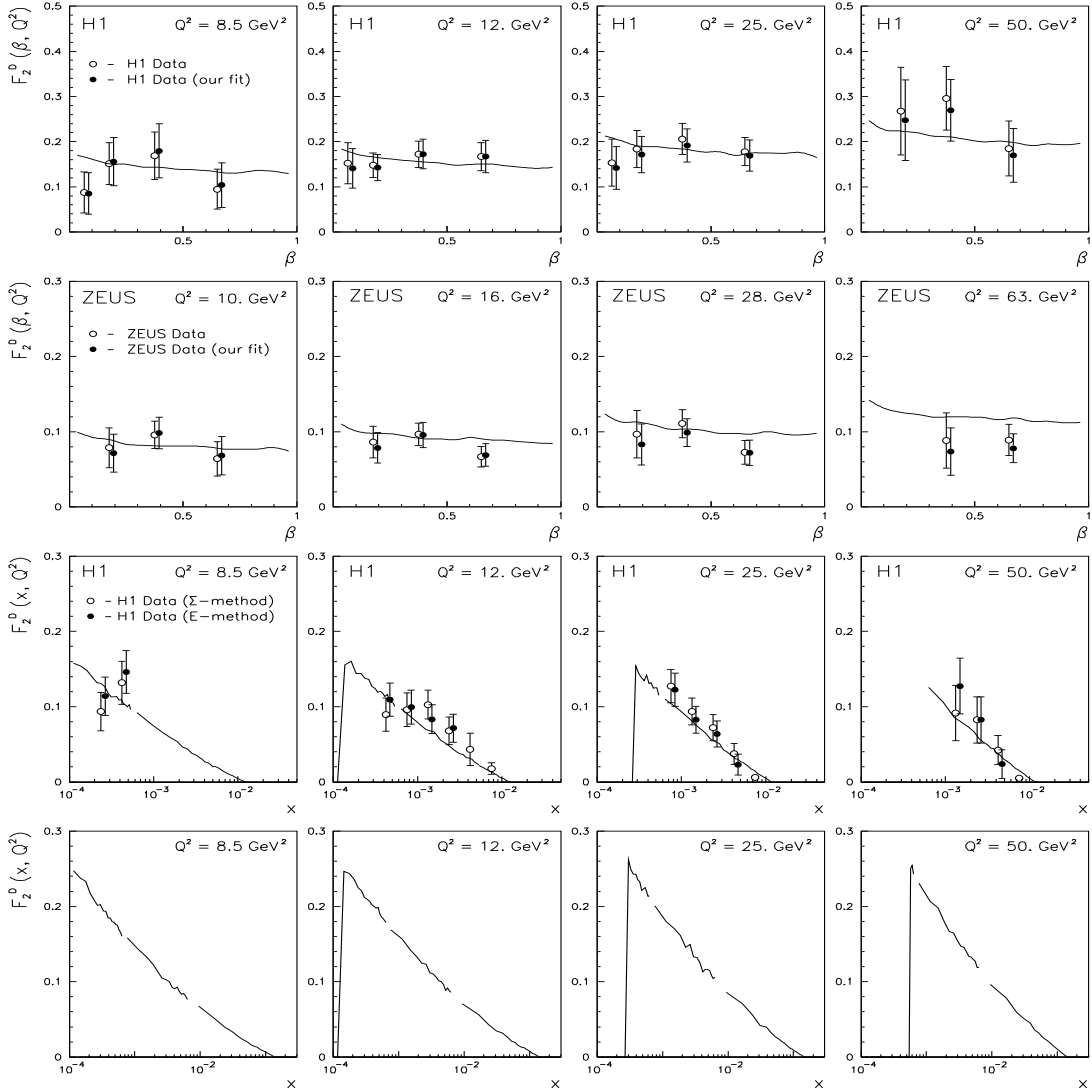,height=17.7cm,width=15.5cm}
\caption
{The double differential structure function $F_2^D$ of ZEUS and H1
in terms of  $\beta$ and $x_{Bj}$, predicted by the model (full line) 
and compared with the experimental data. The black points on 
$\beta$-distribution correspond to $\alpha_{\pom}(0)$ = 1.069, obtained 
in our power-law fit to combined set of HERA data. 
The black and white points on $x_{Bj}$-distribution are observations using 
two different analysis methods [21], the vertical lines are pure 
kinematic cuts. Note that the $15\%$ contribution
due to double diffraction has been subtracted from data [18]. 
The last four pictures show the expected contribution of 
single diffraction dissociation, integrated with the relaxed upper 
limit on $\eta_{\pom}$ (see text above).}
\label{Fig:beta}
\end{figure}

\end{document}